\documentclass[aps,prx,twocolumn,showpacs,superscriptaddress,letterpaper,longbibliography]{revtex4-1}

\usepackage{graphicx}
\usepackage{mdframed}
\usepackage{framed}
\usepackage{indentfirst}
\usepackage{natbib}
\usepackage{amsmath,amssymb}
\usepackage{subfigure}
\usepackage{enumitem}
\usepackage{epstopdf}
\usepackage{xfrac}
\usepackage{multirow}

\begin{document}

\title{Students' views about the nature of experimental physics}


\author{Bethany R. Wilcox}
\affiliation{Department of Physics, University of Colorado, 390 UCB, Boulder, CO 80309}

\author{H. J. Lewandowski}
\affiliation{Department of Physics, University of Colorado, 390 UCB, Boulder, CO 80309}
\affiliation{JILA, National Institute of Standards and Technology and University of Colorado, Boulder, CO 80309}

\begin{abstract}
The physics community explores and explains the physical world through a blend of theoretical and experimental studies. The future of physics as a discipline depends on training of students in both the theoretical and experimental aspects of the field.  However, while student learning within lecture courses has been the subject of extensive research, lab courses remain relatively under-studied.  In particular, there is little, if any, data available that addresses the effectiveness of physics lab courses at encouraging students to recognize the nature and importance of experimental physics within the discipline as a whole.  To address this gap, we present the first large-scale, national study ($N_{institutions}=75$ and $N_{students}=7167$) of undergraduate physics lab courses through analysis of students' responses to a research-validated assessment designed to investigate students' beliefs about the nature of experimental physics.  We find that students often enter and leave physics lab courses with ideas about experimental physics as practiced in their courses that are inconsistent with the views of practicing experimental physicists, and this trend holds at both the introductory and upper-division levels.  Despite this inconsistency, we find that both introductory and upper-division students are able to accurately predict the expert-like response even in cases where their views about experimentation in their lab courses disagree.  These finding have implications for the recruitment, retention, and adequate preparation of students in physics.  
\end{abstract}

\maketitle

\section{\label{sec:intro}Introduction}
The discipline of physics is built on the interplay of theory and experiment. Theory helps to give meaning to the results of experiments and guides new experimental directions. In turn, experimental measurements test predictions of theoretical models and help to refine these models to push the frontiers of physics knowledge. It is impossible to truly understand physics without understanding the role of experimentation in building and supporting the body of physics knowledge. Undergraduate physics education programs acknowledge the importance of experimentation and require students to engage in the activity through instructional lab courses and undergraduate research. However, unlike lecture courses on physics theory, student outcomes from lab courses remain largely unexplored by education researchers. As we work to better prepare our students for graduate school or future careers in our increasingly science- and technology-based world, we must better understand student learning within these experimental learning environments.

The physics education research (PER) community has, until recently, concentrated its efforts on understanding and improving undergraduate education primarily in introductory lecture courses (see Refs.\ \cite{docktor2010dber,mcdermott1999resource} for reviews). However, over the last decade, PER researchers have expanded their studies into upper-division courses (e.g., \cite{passante2015quantum, wilcox2015sov, smith2015thermo, meltzer2012resource}), and, most importantly for the current work, into the laboratory domain \cite{zwickl2015modeling, holmes2015pnas, devore2016amplifiers}. Thus, investigations of student learning in lab courses represent a frontier subfield of PER. In the work described herein, we contribute to the fundamental knowledge in this field using a national-scale study to evaluate particular dimensions of student success in laboratory physics courses.

With respect to students' success in lab courses, physics faculty members often identify a large number of possible goals for these courses including development of lab skills (e.g., experimental design, data analysis, scientific communication, and modeling of experiments) and understanding the nature and process of experimental physics \cite{zwickl2013adlab}. These goals are echoed by guidelines from professional physics societies and other national calls \cite{AAPT2015guidelines, feynman1998goals, nap2013nrc, nrc2003bio, olson2012excel}. Here, we concentrate on the goal of having students develop expert-like views and beliefs about the nature and process of experimental physics. 

Previous work in lecture-based physics courses suggests that typical courses are not accomplishing the goal of improving students' ideas about the nature and importance of physics more generally \cite{madsen2015meta}. For example, surveys of students' attitudes and beliefs about physics typically show a shift to more novice views after instruction \cite{adams2006class, redish1998mpex}.  Prior work has also demonstrated that students beliefs about the nature of physics, and science generally, are correlated with both their self-reported interest in physics \cite{perkins2005class, perkins2006class} and their performance on assessments of their conceptual understanding \cite{perkins2004class}.   As both interest and performance are important aspects of a students' persistence in a given major, these findings have implications for the recruitment and retention of students in the physics major.

\begin{figure}[b]
\begin{minipage}{0.95\linewidth}
   \vspace{1mm}\flushleft {\bf If I don't have clear directions for analyzing data, I am not sure how to choose an appropriate analysis method.} \\
   \hspace{31mm}Strongly  \hspace{23mm} Strongly\\
   \hspace{31mm}disagree \hspace{2mm}1 \hspace{2mm}2 \hspace{2mm}3 \hspace{2mm}4 \hspace{2mm}5 \hspace{3mm}agree \\
   \flushleft
   \begin{minipage}{0.515\linewidth}
      \flushleft\emph{What do YOU think when doing experiments for class?}\vspace{2mm} \\
      \emph{What would experimental physicists say about their research?}
   \end{minipage}
   \begin{minipage}{0.3\linewidth}
      \includegraphics[width=22mm]{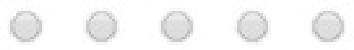}\vspace{5mm}
      \includegraphics[width=22mm]{radio.eps}\vspace{4mm}
   \end{minipage}
\end{minipage}
\caption{An example item from the E-CLASS.  Students are asked to rate their agreement with the statement from their own perspective and that of an experimental physicist.  See Fig.\ \ref{fig:byItem} or Ref.\ \cite{ECLASSwebsite} for a list of all item prompts. }\label{fig:example}
\end{figure}

Extensions of this type of work to students' ideas about experimental physics, however, are less common. Beyond the work described here, there have been no large-scale investigations that characterize students' views about the nature and importance of experimental physics as practiced in their lab courses or how these views compare to those of practicing physicists.  This paper addresses this gap using a large data set of student responses to the research-based, laboratory assessment known as E-CLASS (Colorado Learning Attitudes about Science Survey for Experimental Physics) \cite{zwickl2014eclass,wilcox2016eclass}.  The E-CLASS is a research-based and validated survey that probes students' views about the nature and importance of experimental physics.  In the E-CLASS, students are asked to rate their level of agreement to 30 statements, such as, ``I am usually able to complete an experiment without understanding the equations and physics ideas that describe the system I am investigating.''  Students rate their level of agreement -- from strongly agree to strongly disagree -- to each statement both from their own perspective when doing experiments in their laboratory course, and from the perspective of a hypothetical experimental physicist (Fig.\ \ref{fig:example}).  

Over the past seven semesters, we have collected pre- and postinstruction responses to the E-CLASS from more than 7000 students from 130 distinct physics lab courses spanning 75 different institutions.  Several of these institutions administered E-CLASS in multiple semesters of the same course during data collection. Thus, the full data set includes matched responses from 206 separate instances of the E-CLASS.  This national data set includes both introductory and upper-division courses, and a variety of different institution types. We have previously presented the details of how the E-CLASS was developed, validated, and administered online in order to aggregate such a large database of students' responses \cite{zwickl2014eclass, wilcox2016eclass, wilcox2016admin}. In addition, responses to the survey have allowed us to explore the role gender plays in performance on E-CLASS \cite{wilcox2016gender}. We have also measured the impact of different types of lab activities on E-CLASS scores, where we found students in courses that included at least some open-ended activities outperformed students in courses with only guided labs \cite{wilcox2016structure}. Similarly, we have measured a significant improvement in E-CLASS scores in courses that use well-established transformed curricula compared to traditional labs at the introductory level, and the increase is significantly larger for women \cite{wilcox2016pedagogy}. Finally, we have shown that courses that focus more on developing lab skills outperform courses that focus more on reinforcing physics concepts, and again, the increase is significantly larger for women \cite{wilcox2016focus}. 

Our previous work has focused almost exclusively on what factors (e.g., student gender, instructional approach, etc.) impact students' performance on the E-CLASS.  In the present work, we take a broader view of what we can learn about students' ideas and views about experimental physics from this extensive data set of students' responses to this assessment tool. In doing so, we address the following research questions.

\textit{What are students' views on the nature of experimental physics as practiced in their lab courses, and how do these views shift after laboratory instruction?}

\textit{Do students' views on the nature of experimental physics as practiced in their lab courses vary based on the level of the course?}

\textit{What do students think that expert physicists believe about the nature of experimental physics as practiced by experimental physicists?}

\textit{How do students' views of experimental physics in their courses differ from their predictions of experts' views?}

The answers to these questions provide a snapshot of the status of students' ideas and the effectiveness of undergraduate physics curricula at aligning these ideas with those of practicing physicists.  Moreover, in answering these questions, a major goal of this paper is to provide a comprehensive summary of the national data set that lab instructors can use as a reference when examining their own students' performance on the E-CLASS.  Overall, we find that students both enter and leave undergraduate courses with a variety of ideas about the nature and importance of experimental physics as practiced in their courses, and that some of these views are inconsistent with the views of experts.  We also find that these views tend not to shift significantly over the course of one semester of laboratory instruction.  Our findings provide a valuable resource for instructors and curriculum developers interested in helping students to understand and appreciate the experimental nature of physics as a discipline.  These results also have implications for instructors and researchers interested in the recruitment and retention of students into the physics major, as well as the development of a scientifically-literate citizenry who are capable of taking an informed stance on the importance of science and technology within our society.

\section{\label{sec:methods}Methods}

In this section, we discuss the assessment instrument, data sources, student and institution demographics, and analysis methods used for this study.  

\subsection{\label{sec:validation}Instrument validation}

The E-CLASS is a research-based and validated assessment instrument \cite{wilcox2016eclass}.  After its initial development, the E-CLASS was reviewed by 23 practicing experimental physicists \cite{zwickl2014eclass}.  These expert responses both ensured that the E-CLASS prompts were clear and valuable, and established the consensus expert-like responses to each item.  Twenty-four of the E-CLASS items had greater than 90\% agreement from our pool of experts.  The remaining six questions all had greater than 70\% agreement; additional discussion of these six questions and motivation for retaining them is presented in Ref.\ \cite{zwickl2014eclass}.  The E-CLASS was also given to 42 students in an interview setting, in which students responded to each prompt while talking through the reasoning behind their selection \cite{zwickl2014eclass}.  These student interviews ensured that students were consistently interpreting the prompts and responding in ways that were consistent with their articulated reasoning.  

\begin{table*}
\caption{Demographic breakdown of the full data set for both FY (first-year) and BFY (beyond-first-year) courses.  Number (N) of courses refers to the number of distinct courses, and percentages represent the percentage of students rather than the percentage of courses.  For Major and Gender demographics, the totals may not sum to 100\% as some students did not complete these questions or selected `Other' as their gender.     }\label{tab:dems}
\begin{ruledtabular}
   \begin{tabular}{ l c c c c c c c c c c}
       & \multicolumn{2}{c}{N} & & \multicolumn{2}{c}{Gender} & & \multicolumn{4}{c}{Major} \\
       & Courses & Students & & Women & Men & & Physics & Engineering & Other Science & Non-Science  \\
     \hline
     FY & 63 & 5609 & & 44\% & 54\%  & & 7\% & 27\% & 56\% & 9\%  \\
     BFY & 67 & 1558 & & 19\% & 78\% & & 71\% & 19\% & 7\% & 1\%  \\
   \end{tabular}
\end{ruledtabular}
\end{table*}  

The E-CLASS was also extensively tested for statistical validity and reliability using a subset of the current data set (see Sec.\ \ref{sec:data}), which included $N=3591$ student responses from 71 distinct courses at 44 institutions \cite{wilcox2016eclass}.  These data were used to demonstrate that, in addition to having test and item scores within acceptable ranges \cite{ding2009mcAnalysis}, E-CLASS scores were stable against retesting effects (i.e., test-retest reliability); independent of how long students took to complete the assessment (i.e., time-to-completion reliability); and independent of whether students completed the assessment in-class or out-of-class (i.e., testing environment reliability).  E-CLASS scores also adequately distinguished between high and low performing students (i.e., whole-test and item discrimination \cite{ding2009mcAnalysis}), and provided consistent scores on subsets of items (i.e., internal consistency \cite{cortina1993ca}).  An exploratory factor analysis \cite{oRourke2013factor} on these data also showed that, in accordance with its initial design, the E-CLASS items did not factor into coherent subgroups of related questions whose scores could be reported in aggregate, rather than as individual items \cite{wilcox2016eclass}.

\subsection{\label{sec:data}Data sources}

Data presented here are composed of students' responses to the E-CLASS, which includes items designed to measure students' beliefs about the nature and importance of experimental physics, as well as their confidence when performing physics experiments.  The full list of E-CLASS prompts is given in Fig.\ \ref{fig:byItem} and is also available from Ref.\ \cite{ECLASSwebsite}.  The E-CLASS items were developed to target a wide range of learning goals in order to make the assessment relevant for both introductory and advanced laboratory courses.  

Data for this study were collected over the course of seven semesters using a centralized, online system \cite{wilcox2016admin}.  The data set includes student responses from 130 distinct courses from 75 different institutions.  Institutions in the data set span a range of institution types including 2-year ($N=3$) and 4-year colleges ($N=36$), as well as masters ($N=8$) and Ph.D. granting universities ($N=28$).  Additionally, in some courses, the E-CLASS was administered during multiple semesters of the same course, thus, the full data set includes matched responses from 206 separate instances of E-CLASS.  These courses include both first-year (FY), introductory courses as well as more advanced, beyond-first-year (BFY) courses.  Only students who completed both the pre- and postinstruction E-CLASS were included in the final data set ($N=7167$).  Table \ref{tab:dems} reports the demographic breakdown of these students with respect to course level, gender, and major.  

\subsection{\label{sec:analysis}Scoring and analysis}

For each of the questions on the E-CLASS, students were presented with five possible response options, from ``Strongly disagree'' to ``Strongly agree'' (see Fig.\ \ref{fig:example}).  For the purposes of scoring, students' responses were classified simply as agree, disagree, or neutral by collapsing ``Strongly (dis)agree'' and ``(dis)agree'' into a single category.  Students were then awarded 1 point if their response was consistent with the established, expert-like response, and 0 points otherwise.  The accepted expert-like response was established during the development of the E-CLASS based on responses from physics laboratory instructors and practicing experimental physicists \cite{zwickl2014eclass}.  Using this scoring scheme, the average of all students scores on a particular item represents the fraction of students who responded favorably (i.e., consistent with experts) to that item.  We also calculate an overall E-CLASS score for each student, which is given by the average of that student's individual item scores.  Thus, a student's overall score represents the fraction of the 30 E-CLASS items for which that student gave a favorable response.  Recall that students provide two responses for each of the 30 E-CLASS prompts -- one representing their perspective regarding experimentation in their lab courses and one representing their prediction of what an experimental physicist might say about their research (Fig.\ \ref{fig:example}).  Individual item and overall scores are calculated separately for these two sets of responses.  

Note, the 2-point scoring scheme described above is different than the 3-point scoring scheme that has been used in the majority of the prior E-CLASS publications \cite{wilcox2016eclass, wilcox2016gender, wilcox2016focus}.  However, the 2-point scheme is consistent with the representations used in the reports received by instructors using E-CLASS through our centralized system \cite{wilcox2016admin}.  Thus, in order that the data and analysis reported here can provide a comprehensive reference for these instructors, we have opted to use the 2-point scoring scheme here.

Section \ref{sec:results} reports means of students' overall and by-item scores for students in FY and BFY courses.  To determine the statistical significance of differences between various score distributions, we use the non-parametric Mann-Whitney U test \cite{mann1947mwu}.  In cases where we calculate multiple comparisons, we utilize Holm-Bonferroni corrected p-values to account for multiple-testing effects \cite{holm1979hb}.  Additionally, we characterize the size of the statistically significant differences in scores using Cohen's $d$ \cite{cohen1988d}.  Due to the large size of the matched data set, some of the statistically significant differences we observed fell well below the threshold of what is generally considered a small effect ($d=0.2$ \cite{cohen1988d}).  However, small effects can be practically significant, particularly when many small effects in the same direction combine to form a larger overall effect.  For the current analysis, we are interested in identifying the more dominant trends in students' responses, including looking at small effects on individual items.  To balance this with the large statistical power of our data set, we set a threshold for practical significance of $d>0.1$ that preserves some of the small effects while eliminating effects that are simply too small to warrant specific attention.  Thus, when comparing means in the following sections, we will distinguish between results that were only statistically significant and those that were both statistically and practically significant ($p<0.05$ and $d>0.1$).

\section{\label{sec:results}Results}

This section presents findings with respect to students' responses to the E-CLASS from their perspective and from the perspective of a hypothetical experimental physicist, as well as how the two compare.

\subsection{\label{sec:resultsPersonal}What do students believe?}

\begin{figure*}
\includegraphics[width=0.9\linewidth]{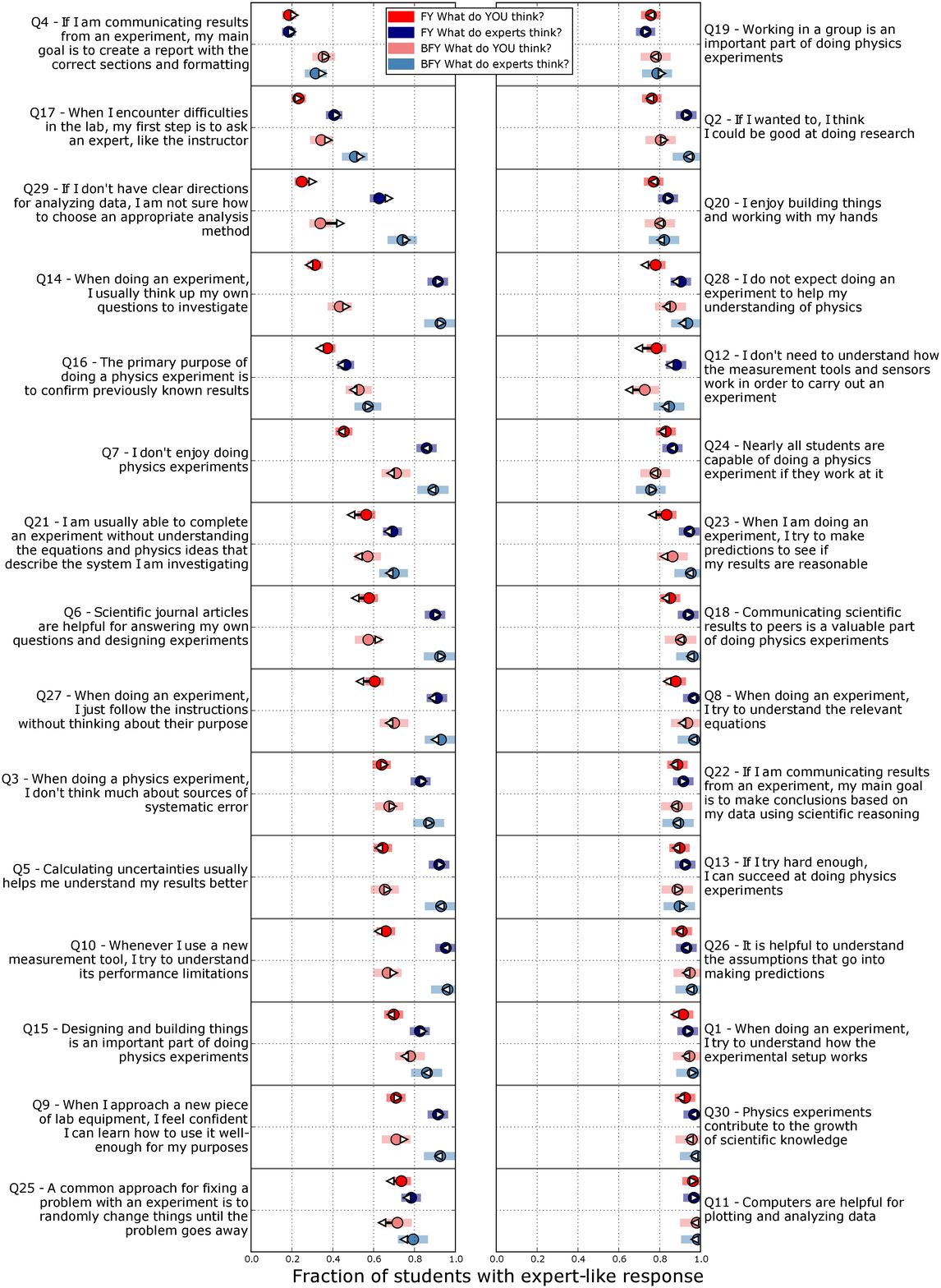}
\caption{Fraction of students with expert-like responses for each E-CLASS item.  Items are ordered by fraction favorable among FY students.  Solid circles indicate the preinstruction fraction while the arrow indicates the postinstruction fraction and points in the direction of the shift from pre- to postinstruction.  Shaded bars indicate the 95\% confidence interval on the preinstruction fraction.  Red and pink represent students' responses from their own perspective for FY and BFY students respectively, and Blue and teal points represent students' predictions of what an experimental physicist might say for FY and BFY students respectively (color online).  }\label{fig:byItem}
\end{figure*}

Here, we examine students' responses to each of the E-CLASS prompts with respect to their views about experimentation in their lab courses (see Fig.\ \ref{fig:example}).  To explore general trends in students' responses to the E-CLASS prompt targeting what they think, we first examine the average overall score (i.e., fraction of items with favorable responses) both before and after instruction.  Table \ref{tab:overall} reports pre- and postinstruction overall E-CLASS scores for both FY and BFY courses.  The motivation for separating the data by course level rather than reporting the aggregate statistics is twofold.  First, prior analysis of E-CLASS data shows that trends in the responses from students in FY and BFY courses are meaningfully different \cite{wilcox2016eclass, wilcox2016gender, wilcox2016structure}.  Second, due to the large number of students in the FY courses (see Table \ref{tab:overall}), the aggregate statistics are strongly driven by the FY population.  In addition to examining students' overall scores, we also look at students' responses to the E-CLASS items individually.  Fig. \ref{fig:byItem} presents the pre- and postinstruction fraction of students who responded favorably to each of the 30 E-CLASS items.  

\begin{table}[b]
\caption{Average (standard error of the mean) of the fraction of items answered favorably on both the pre- and post-tests for students in FY and BFY courses for the questions targeting students personal views when doing experiments for class.}\label{tab:overall}
\begin{ruledtabular}
   \begin{tabular}{ l c c c }
      & N &Pre& Post \\
      \hline
      FY & 5609 & 0.679 (0.002) & 0.655 (0.002) \\
      BFY & 1558 & 0.727 (0.003) & 0.723 (0.004) \\
   \end{tabular}
\end{ruledtabular}
\end{table}

Table \ref{tab:overall} shows that in the FY courses, students, on average, responded consistently with experts on roughly two-thirds of the E-CLASS items when asked about their views on experimental physics as practiced in their lab courses.  This fraction favorable decreased significantly over the course of one semester (or quarter) of instruction ($p\ll0.01$, $d=-0.15$).  Rather than being driven by particularly large shifts on a few items, this overall negative shift in the FY was driven by small, but statistically significant, negative shifts across 17 of the E-CLASS items.  Alternatively, Table \ref{tab:overall} shows that students in the BFY courses responded favorably to on average just under three-quarters of the E-CLASS items, and this fraction did not shift significantly over one semester (or quarter) of instruction ($p=0.6$).  

These results suggests that both FY and BFY courses had, at most, only a small impact on the views students expressed with respect to the overall E-CLASS score or an individual item; however, in FY courses, the cumulative effect tended to drive students towards less expert-like views.  This finding is consistent with results from similar research-based assessments used in lecture courses, which also find that students perceptions of physics or science more generally, typically become less expert-like over the course of an introductory physics course \cite{adams2006class, redish1998mpex}.  

In addition to looking at students' pre- to postinstruction shifts to explore the impact of laboratory instruction, we also examine students' raw postinstruction scores to determine which E-CLASS items elicit the least (or most) expert-like responses.  Insight into areas in which students' views about the nature of experimental physics as practiced in their lab courses are least aligned with experts can provide a guide to help focus the efforts of instructors interested in improving students' beliefs.  

Four items on the E-CLASS resulted in less than 50\% of students providing favorable responses in both FY and BFY courses (see Fig. \ref{fig:byItem}).  These items are given below, along with the established expert-like response in parentheses.  

\begin{itemize}[noitemsep, topsep=0pt]
  \item Q4 - If I am communicating my results from an experiment, my main goal is to have the correct sections and formatting (disagree)
  \item Q17 - When I encounter difficulties in the lab, my first step is to ask an expert, like the instructor (disagree)
  \item Q14 - When doing an experiment, I usually think up my own questions to investigate (agree)
  \item Q29 - If I don't have clear directions for analyzing data, I am not sure how to choose an appropriate analysis method (disagree).
\end{itemize}

\noindent The latter three of these questions represent all of the questions on the E-CLASS related to what can be loosely described as student autonomy, or their ability to direct an experiment, overcome difficulties, and select analysis methods without guidance from an authority figure.  It is worth acknowledging that low performance from first years on these questions is not irrational nor unexpected given the often rushed nature and content (rather than skills) focus of many traditional FY labs.  However, low performance on these questions in the BFY courses as well is particularly concerning given that the ability to work autonomously is an important characteristic of successful graduate students and professional physicists.  

For first year courses only, a further three questions resulted in less than 50\% of students providing favorable postinstruction responses in the (see Fig.\ \ref{fig:byItem}):

\begin{itemize}[noitemsep, topsep=0pt]
  \item Q16 - The primary purpose of doing physics experiments is to confirm previously known results (disagree)
  \item Q7 - I don't enjoy doing physics experiments (disagree)
  \item Q21 - I am usually able to complete an experiment without understanding the equations and physics ideas that describe the system I am investigating (disagree).
\end{itemize}

\noindent Given that the majority of students in FY physics lab courses are not physics majors, the low score on Q7 may have been driven in part by the fact that, for most of these students, the physics lab is a required course not directly related to their chosen major.  Scores on this question did not shift significantly from pre- to postinstruction.  Scores on Q21, however, did show a statistically significant negative shift.  This result is potentially surprising given that in roughly 80\% of the courses in our data set, the instructor reported ``reinforcing physics concepts'' as one of the main goals of their lab course.  

Alternatively, Fig.\ \ref{fig:byItem} shows that many of the E-CLASS prompts elicited expert-like postinstruction responses from more than 80\% of the students in both FY and BFY courses.  For all of these questions, the majority of students both start and end the course with views consistent with those of practicing physicists.  One of these high performing items for both FY and BFY students was Q30 (``Physics experiments contribute to the growth of scientific knowledge'').   At first glance, high performance from FY students on Q30 seems potentially at odds with the fact that more than 50\% of these students' agreed with the statement that ``The primary purpose of physics experiments is to confirm previously known results'' (Q16).  One possible explanation for this apparent contradiction is that these students consider results that are predicted by theory or consistent with an established mathematical model as ``previously known results.''  Thus, while the vast majority of experiments have an expected or predicted result, students may still see this as contributing to the growth of scientific knowledge because these predictions must be tested empirically.

\subsection{\label{sec:resultsLevel}How do students' beliefs vary by course level?}

In addition to the general differences shown in the previous section, there was a statistically significant difference in the overall E-CLASS scores of FY and BFY students, with BFY students scoring higher (see Table \ref{tab:overall}).  Additionally, 24 of the 30 E-CLASS items showed a statistically significant difference between the score distributions from students in different level courses.  Of these items, FY students scored higher both statistically and practically ($d>0.1$) in only one case --- Q24, ``Nearly all students are capable of doing physics experiments if they work at it.''  There are a number of possible factors that may contribute to lower performance on this item from BFY students, including increased difficulty characteristic of upper-level experiments, and the significant attrition typically observed between FY and BFY courses.  Thus, not only do BFY students encounter more complex and challenging experiments, but they are a highly self-selected group that has seen more of their classmates fail or choose to leave the physics major.  

The size of the difference between FY and BFY students varied significantly by item; however, the five questions with the largest differences ($d>0.3$) between FY and BFY scores (Q4, Q17, Q14, Q16, Q7), were also five of the six lowest scoring items among the FY population (see Fig.\ \ref{fig:byItem}).  This trend suggests that these items, in addition to eliciting the smallest fractions of favorable responses, also provided the greatest discrimination between courses of differing levels.  

While, these findings clearly indicate a significant difference in students' responses to the E-CLASS based on the level of their course, this analysis does not address the causal mechanism for this trend.  Higher scores in BFY courses could be due to: the cumulative impact of instruction as students progress through the curriculum; a selection effect based on which students enter into, and persist in, a STEM or physics program; or a combination of these and/or other factors.

\subsection{\label{sec:resultsExpert}What do students think experts believe about experimental physics?}

To this point, we have discussed students' responses to only the first of the two E-CLASS questions (see Fig.\ \ref{fig:example}), in which students are asked to rate their agreement to each statement based on what they think when completing experiments in class.  Next, we examine their responses to the second question asking them to predict what an experimental physicist might say about their research.  This question, in essence, asks the students what they think the expert-like response would be for each item.  Scoring this set of questions relative to the consensus expert-like response, as before, we examine the average overall score for students' predictions of expert responses (see Table \ref{tab:overallE}).  Table \ref{tab:overallE} indicates that in both FY and BFY courses, students correctly predicted the expert-like response for more than 80\% of the E-CLASS items.  

\begin{table}[b]
\caption{Average (standard error) of the fraction of items answered favorably on both the pre- and post-tests for students in FY and BFY courses for the question targeting students predictions of expert-like response.  }\label{tab:overallE}
\begin{ruledtabular}
   \begin{tabular}{ l c c c }
      & N &Pre& Post \\
      \hline
      FY & 5609 & 0.828 (0.002) & 0.828 (0.002) \\
      BFY & 1558 & 0.850 (0.003) & 0.847 (0.003) \\
   \end{tabular}
\end{ruledtabular}
\end{table}

Fig.\ \ref{fig:byItem} also shows the fraction of students who gave accurate predictions of the expert-like response for each of the 30 E-CLASS items.  Only one item elicited favorable responses from less than 50\% of both FY and BFY students -- Q4, ``If I am communicating the results from an experiment, my main goal is to have the correct sections and formatting.''  This item also resulted in less than 50\% favorable responses with respect to students' own views about experimentation as practiced in their lab courses.  While a larger fraction of unfavorable responses to the personal version of this item is consistent with the grading practices typical of many lab courses, the implication that students also believed that structure and formatting is the major focus of experts is perhaps surprising.  However, while the established expert-like response to this item is ``disagree'' as formatting is clearly not the main goal of these documents, it is worth noting that adherence to style and structure guidelines when writing grant applications and/or publications is a necessary task for a successful physicist.  

An additional two E-CLASS items resulted in less than 50\% of FY students providing accurate predictions of the expert-like responses.  

\begin{itemize}[noitemsep, topsep=0pt]
  \item Q17 - When I encounter difficulties in the lab, my first step is to ask an expert, like the instructor
  \item Q16 - The primary purpose of physics experiments is to confirm previously known results.  
\end{itemize}

\noindent Students' prediction that experts would agree with Q16 is consistent with our earlier hypothesis that these students may be including theory in what they consider to be 'previously known results.'  If this is the case, it suggests that students have difficulty with respect to how theory often involves modeling physical systems, as well as recognizing the role of these models in physics experiments.  

On the other hand, for 22 of the E-CLASS items, 80\% or more of both FY and BFY students were able to accurately predict the expert-like response (see Fig.\ \ref{fig:byItem}).  This finding suggests that even FY students, who may or may not have had any prior experience with experimental physics, had a reasonably good sense of what the 'expert-like' responses were.  The next section looks at the comparison between students' predictions of the expert-like response and their personal views for each of the E-CLASS items.  

\subsection{\label{sec:resultsCompare}How do students' expert predictions and their views compare?}

The previous sections demonstrated that students beliefs about experimental physics as practiced in their lab courses often differ from those of practicing experimentalists (Sec.\ \ref{sec:resultsPersonal}), but that they are fairly accurate when asked to predict the accepted, expert-like response (Sec.\ \ref{sec:resultsExpert}).  Consistent with this result, comparison of the overall E-CLASS scores for students' beliefs (Table \ref{tab:overall}) and their predictions of experts' beliefs (Table \ref{tab:overallE}) show a large ($d=0.8$ and $d=1.1$ for BFY and FY respectively) and statistically significant difference.  In terms of individual items, the difference between the distribution of students' scores on these two sets of prompts was both statistically and practically significant ($d>0.1$) for 27 items for FY students and 22 items for BFY students.  In all cases, students' expert predictions averaged higher than their beliefs about experimentation in their lab courses.  This difference was particularly large ($d>0.5$) in both the FY and BFY populations for 7 items (Q14, Q29, Q7, Q6, Q27, Q5, Q10, see Fig.\ \ref{fig:byItem}).  

These results, combined with the results from the previous section, suggest that students in both FY and BFY courses are good at predicting the expert-like responses, even in cases where their views about experimentation as practiced in their lab courses differ.  This difference between students views and their views of experts also indicates that the students are responding honestly to the prompt targeting their own beliefs within the context of their lab courses, rather than giving the answer they believe is ``correct.''  This finding that students can hold seemingly contradictory beliefs with respect to knowing and learning science has been observed previously \cite{gray2008class,lising2005epistemology}.  Investigating why students form and maintain these contradictory ideas would require the collection of additional qualitative data (e.g., student interviews) and is beyond the scope of this work.  However, previous research suggests that factors that may contribute include students' perception that they are simply different than experimental physicists and thus engage differently in the process of experimental physics \cite{lising2005epistemology}, and/or that the activities in their lab courses are inauthentic and thus do not reflect the actual practice of experimental physics \cite{gray2008class}.

\section{\label{sec:discussion}Summary and Conclusions}

Physics is both an experimental and theoretical science.  Students' exposure to the theoretical grounding of physics comes primarily through numerous core lecture courses taken throughout their undergraduate careers; however, direct experiences with experimental physics are often limited to a few laboratory courses, or, for a subset, undergraduate research experiences.  Helping students to understand the role of experimentation in building and supporting the body of physics knowledge is a critical goal of undergraduate physics programs striving to recruit and retain physics graduates who are prepared for, and interested in, graduate school, industry careers, or simply joining a scientifically-literate citizenry \cite{jtupp2016report}.  This work contributes to a new, but growing, body of literature investigating the status and success of physics lab courses with respect to achieving various learning goals.  Specifically, we present analysis of a large, national data set of student responses to a laboratory focused assessment -- the E-CLASS -- with respect to students' beliefs about the nature and process of experimental physics as practiced in their lab courses, as well as their predictions of what they think experts believe about their research.  

Our findings suggest that undergraduate students in physics lab courses often enter and leave these courses with some ideas about the nature and importance of experimental physics that are inconsistent with those of practicing physicists.  This trend held for both introductory and more advanced students, though upper-level students' views were somewhat more consistent with those of experts than their introductory counterparts.  With respect to the impact of laboratory instruction, participation in an upper-level lab courses did not tend to result in significant shifts in students' views over the course of a single semester, while participation in introductory lab courses tended to result in small negative shifts.  Together, these findings suggest that laboratory physics courses have not been particularly effective at encouraging more expert-like beliefs in students, and in some cases, they have actually resulted in more novice-like beliefs.  We also found that students had a relatively good sense of what the `expert-like' responses are.  Even before instruction and in the introductory courses, students were able to accurately predict the views of practicing physicist on the majority of the E-CLASS items even in cases where their views about experimentatal physics as practiced in their lab courses disagreed.  

There are several important limitations to the work described here.  While our data set is extensive and drawn from a large number of courses and institutions, it is neither comprehensive nor randomly-selected.  In the majority of the courses in our data set, the instructor chose to use the E-CLASS without external pressure from our research group, their department, or their colleagues.  Thus, these instructors represent a self-selected group and may not be representative of the broader population of physics faculty.  Additionally, relatively few 2-year colleges have used the E-CLASS to date, suggesting our results may be dominated by trends in the 4-year college and Ph.D. granting institutions.  

An important caveat to consider when interpreting our results is that improving students' beliefs about the nature and importance of experimental physics is only one of multiple important learning goals for undergraduate physics courses.  Given the variety of potential goals and limited class time, it is nearly always necessary for laboratory instructors to select a subset of these goals to highlight in any given course.  However, for courses in which promoting expert-like attitudes and beliefs is a goal, the E-CLASS can serve as an easy-to-use, research-validated assessment that can help inform efforts to improve these courses.  For example, given the relatively small impact of individual courses, our results suggest that achieving significant improvements in students' views about experimental physics may require larger-scale, programatic initiatives rather than the isolated course-by-course initiatives typical of standard course transformation efforts.  

The data presented here can also serve as an important comparison point by which instructors and researchers can determine the effectiveness of new curricular approaches and pedagogical techniques relative to national trends.  For example, prior work with these data investigated the impact of different types of lab activities on E-CLASS scores, demonstrating that students in courses that include at least some open-ended activities tend to outperform students in courses with only traditional guided labs \cite{wilcox2016structure}. Significant improvement in E-CLASS scores was also documented in courses that use well-established transformed curricula compared to traditional labs at the introductory level, and the increase was significantly larger for women \cite{wilcox2016pedagogy}. Additional investigations showed that courses that focus more on developing lab skills outperform courses that focus more on reinforcing physics concepts within the lab component, and again, the increase was significantly larger for women \cite{wilcox2016focus}. This type of baseline and comparison data is a critical first step towards continuing to improve the undergraduate physics curriculum in order to produce engaged and well-prepared physics graduates and scientifically literate citizens.

\begin{acknowledgments}
This work was funded by the NSF-IUSE Grant No. DUE-1432204 and NSF Grant No. PHY-1734006.  Particular thanks to the members of PER group at the University of Colorado Boulder for all their help and feedback.  
\end{acknowledgments}

\bibliography{master-refs-ECLASS-10_16}

\end{document}